# Wi-Fi Direct Based Mobile Ad hoc Network

Jae Hyeck Lee and Myong-Soon Park
Department of Computer and Radio Communications
Engineering, Korea University
Seoul, South Korea
e-mail: leeadra@gmail.com

Sayed Chhattan Shah
Department of Information Communications
Engineering, Hankuk University of Foreign Studies
Seoul, Korea
e-mail: shah@hufs.ac.kr

*Abstract* —A mobile ad hoc network is a wireless network of mobile devices that communicate with one another without any fixed network infrastructure. Existing mobile ad hoc networks are either based on IEEE 802.11 a or b standard. Both standards provide limited bandwidth and therefore are not suitable for data intensive applications such as automated video surveillance. Wi-Fi Direct is a new technology that enables mobile devices to directly communicate with each other without a wireless access point. Compare to existing technologies it provides data rates of up to 250 Mbps which is sufficient for several mobile data intensive applications. Wi-Fi Direct technology, however, has two main limitations. It does not support communication between two client devices in a group and it also does not provide support for multi hop communication.

This paper describes a Wi-Fi Direct based multi hop mobile ad hoc network. More specifically a routing layer has been developed to support communication between Wi-Fi Direct devices in a group and multi hop communication between devices across a group. The proposed system has been implemented on a group of four Wi-Fi Direct enabled Samsung mobile devices.

*Keywords-component; routing protocol; mobile ad hoc network; IoT communication infrastructure, multi hop communication*

## I. INTRODUCTION

A mobile ad hoc network is a wireless network of mobile devices that communicate with each other without any fixed network infrastructure. Mobile ad hoc network serves as a communication backbone for several systems such as mobile ad hoc cloud, IoT, and mobile CPS [1].

Existing mobile ad hoc networks are either based on IEEE 802.11 a or b standard. Both standards provide limited bandwidth and therefore are not suitable for data intensive applications such as automated video surveillance [2].

Wi-Fi Direct is a new technology that enables mobile devices to communicate with each other without a wireless access point. Compare to existing technologies it provides data rates of up to 250 Mbps which is sufficient for several mobile data intensive applications [3].

Wi-Fi Direct technology, however, has two main limitations. It does not support communication between two client devices in a group and it also does not provide support for multi hop communication which is a key requirement of several IoT and mobile cyber physical system applications [10].

This paper describes a Wi-Fi Direct based multi hop mobile ad hoc network. More specifically, a routing layer has been developed to support communication between Wi-Fi Direct devices in a group and multi hop communication between devices across a group. The proposed system can be easily extended to support communication between devices using various communication technologies such as ZigBee and Bluetooth. This would enable numerous IoT applications where devices with heterogeneous communication technologies are integrated in to a single network. In addition, proposed system would provide a communication infrastructure for mobile ad hoc cloud applications [2], [3] where large amount of data are transferred between cloud nodes.

The rest of the paper is organized as follows. Section II explains the concept, features and limitations of Wi-Fi Direct technology. Section III discusses the related work. Proposed system is described in section IV whereas implementation of system is described in section V.

TABLE I. CHARACTERISTICS OF WIRELESS COMMUNICATION STANDARDS

|  | **IEEE 802.11a** | **IEEE 802.11b** | **Wi-Fi Direct** |
|---|---|---|---|
| **Max Range** | 45m | 45m | 200m |
| **Max Data Rate** | 54Mbps | 11Mbps | 250Mbps |

## II. WI-FI DIRECT

Wi-Fi Direct is a recent technology standardized by Wi-Fi Alliance [4], [6]. It enables mobile devices to directly communicate with each other without a wireless access point.

For communication, Wi-Fi Direct devices have to establish a group. Each device in a group has a role as a Group Owner (GO) or Group Client (GC). GO plays a role of access point. Once group is established other devices can join the group as group clients. GO can connect with multiple devices at the same time but GCs can only connect to GO. Communication among devices in a group is depicted in Figure 1. Wi-Fi Direct also support concurrent mode in which devices can communicate via multiple wireless communication technologies. Wi-Fi Direct concurrent mode is depicted in Figure 2.

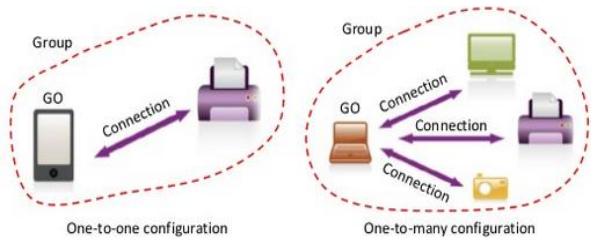
Figure 1. Communications in a group.

Lifetime of a group and communications are related to GO's behavior, because the roles do not change once group is established, and all communications in a group are performed through a GO.

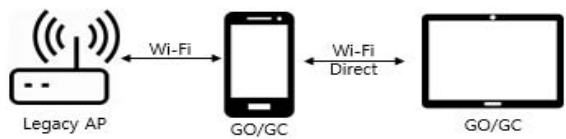
Figure 2. Wi-Fi Direct Concurrent Mode

### A. Device Discovery Process

The objective of device discovery phase is to find devices within a communication range. It consists of two sub-phases: *Scan* and *Find*.

*Scan:* During Scan sub-phase, each device collects information about nearby devices by scanning all supported wireless channels. Scan sub-phase also enables Wi-Fi Direct devices to discover potential legacy devices operating on channels in addition to social channels (1, 6 or 11 in the 2.4 GHz band). The Scan sub-phase has a pre-defined duration and is followed by Find sub-phase [5].

*Find*: During Find sub-phase, device alternates between two states: a *search* state, in which the device performs active scanning by sending Probe Requests; and a *listen* state, in which the device listens for Probe Requests and then responds with Probe Responses. The amount of time that a device spends on each state is randomly distributed, typically between *100ms* and *300ms* [6].

### B. Group Formation Process

There are three different procedures for a group formation: (i) *standard*, in which devices discover each other and then negotiate GO role, (ii) *autonomous*, in which node elects itself as GO and announces its presence through beacon messages, (iii) *persistent*, in which a device declare a group as persistent using an attribute present in beacon frame. This paper focuses on the standard procedure [5].

GO role negotiation involves three-way handshake process: Request-Response-Confirmation. The two devices decide the role of GO and the channels on which group will operate. To decide GO role, devices exchange a numerical parameter, *GO Intent value*, during a three-way handshake process, and the device with highest intent value becomes a GO. To prevent a conflict in a situation where two devices have the same intent value, a *tie-breaker bit* is included in GO Negotiation Request [6].

Once the GO is identified, an authentication procedure is performed based on Wi-Fi Protected Setup in order to establish a secure wireless connection [7]. This is followed by address configuration phase in which devices receive an IP address assigned by DHCP server running on GO [5].

### C. Limitations of Wi-Fi Direct Technology

Wi-Fi Direct technology has two main limitations. It does not support communication between client devices in a group. Group clients in a group cannot communicate with each other. They only can communicate with group owner as depicted in Figure 3-a. Even though group clients GC1 and GC2 are in a communication range of each other but they cannot communicate.

Wi-Fi Direct technology also does not provide support for multi hop communication. As shown in Figure 3-b, group owner GO cannot communicate with out of range device D even though that device is in range of GC which is connected to GO. This is due to the protocol that group client cannot become a group owner in another group and group client cannot establish connections to multiple devices.

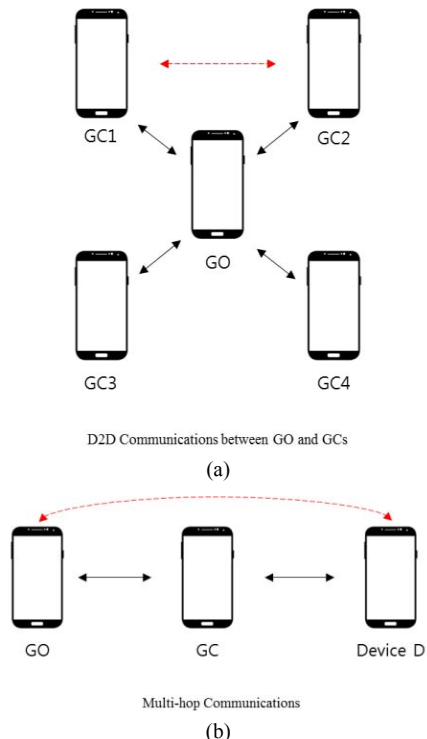
Figure 3. Limitations of Wi-Fi Direct.

### III. RELATED WORK

To overcome the limitations, few schemes such as [7]-[11] have been proposed. For example, authors in [7] proposed an architecture which creates content centric ad hoc network comprising of Wi-Fi Direct enabled smartphones. In the proposed architecture, it is claimed that group clients connected to group owner on channel x in one group can communicate with another group clients on another channel using Wi-Fi Direct concurrent mode. However, according to

117

standard, in Wi-Fi Direct concurrent mode a group client cannot become a group owner in another group [4]. In addition, the proposed system is implemented on network simulator ns-3 which does not support Wi-Fi Direct standard. It provides support for Wi-Fi ad hoc mode [7].

Authors in [8] proposed a tunneling mechanism that allows inter-group communication between Android smartphones. To support multi hop communication a node sends a packet to intermediate node. The packet includes source address, destination address and sequence number. An intermediate node operating in relay-enabled mode then broadcasts the packet. The broadcast of packet degrades communication performance.

Authors in [9] proposed a multi-group networking scheme using [8]. The topology and IP address assignment are same as in [8]. The difference is that [9] uses CCN module in order to send and receive contents. CCN module has two main components: Content Routing Table (CRT) and Pending Interest Table (PIT). CRT stores IP addresses of nodes in a communication range. PIT records information to route content to requester by storing IP address of a node from which a request was received. The schemes proposed in [8] and [9] also do not support communication with a device which is not included in a group.

## IV. PROPOSED SYSTEM

Wi-Fi Direct technology does not support communication between clients in a group and it also does not provide support for multi hop communication.

In this section, we describe Wi-Fi Direct based mobile ad hoc network architecture that aims to provide support for communication between Wi-Fi Direct devices in a group and multi hop communication between Wi-Fi Direct devices across a group. The proposed architecture can be easily extended to support communication between devices using various communication technologies such as ZigBee and Bluetooth. This would enable numerous IoT applications where devices with heterogeneous communication technologies are integrated in to a single network. It will also provide opportunities for robust and efficient data transfers between nodes as depicted below.

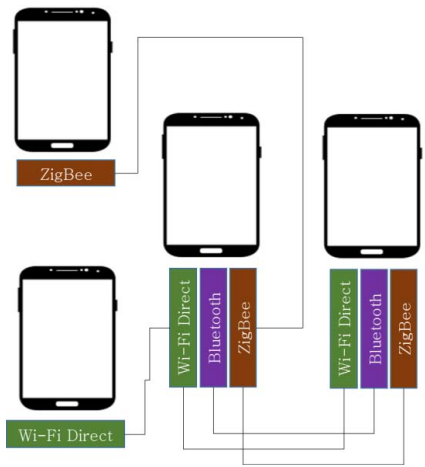

The architecture of proposed system is given in Figure 4. It consists of three layers: Application Layer, Routing Layer, and Android Layer. Application Layer includes applications such as video surveillance and smart home. Routing Layer provides support for single and multi-hop communication whereas Android Layer includes operating system and Wi-Fi Direct interface.

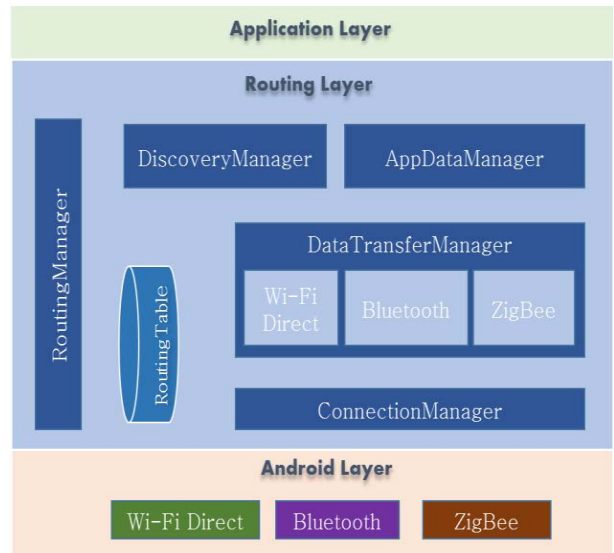

Figure 4. Proposed system architecture.

Components of routing layer are described below while relationship between components is depicted in Figure 5.

### A. DiscoveryManager

DiscoveryManager performs following key tasks. (a) Discover devices in a communication range. (b) Periodically broadcast information stored in routing table to neighbor nodes. (c) Insert or update routing information received during discovery process. Discovery Manager uses Data Transfer Manager to send, receive and broadcast discovery packets. It also communicates to Routing Table to manage routing information.

### B. ConnectionManager

ConnectionManager is responsible for establishing or terminating a connection to a device. It is used by Data Transfer Manager to establish or terminate a connection.

### C. DataTransferManager

DataTransferManager performs following key tasks. (a) Provide an interface to a respective wireless communication technology such as Bluetooth, Wi-Fi Direct and ZigBee. (b) Send and receive all type of packets such as control packets and data packets related to an application.

### D. RoutingManager

RoutingManager selects a route to a destination as per application requirement. For real-time applications it selects a route with minimum latency whereas for non-real-time applications it selects a route which consumes less energy.

RoutingManager interacts with RoutingTable to access routing information, Data Transfer Manager to send data to network or receive data from the network, and AppDataManager to send data to an application or receive data from an application for the transmission across the network.

### E. RoutingTable

It stores routing and quality of service related information of discovered devices.

### F. AppDataManager

AppDataManager hides all the complexity and provides a simple and easy to use interface to an application for transmission of data across the network.

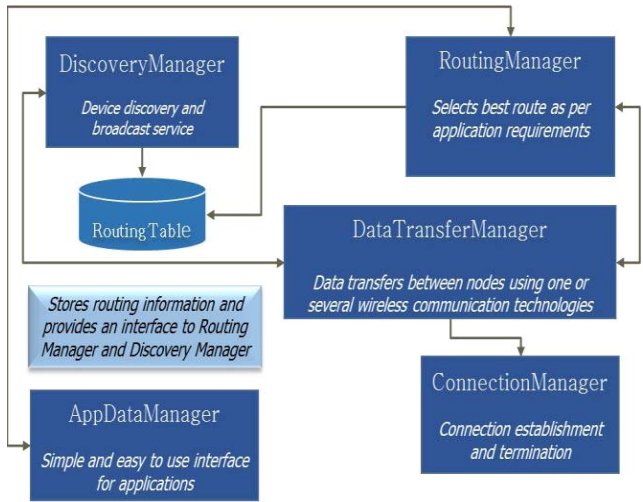

Figure 5. Relationship between routing layer components.

## V. IMPLEMENTATION

For demonstration of proposed system, four Wi-Fi Direct enabled Samsung Galaxy devices were used. At application layer a chatting application module was developed to send text and image data. At Routing Layer, Discovery Manager, Routing Manager, Routing Table, Data Transfer Manager, and AppDataManager were developed using Android Wi-Fi Direct library.

Wi-Fi Direct based mobile ad hoc network formation process began with a discovery process initiated by a node A. During the discovery process, a discovery request packet is broadcast. Status of routing table at each node after discovery process is given in Figure 6.

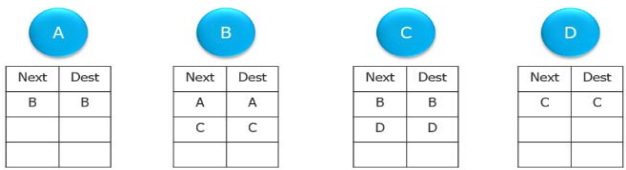

Figure 6. Status of routing tables after discovery process.

After discovery of nodes in a communication range, each node periodically broadcast a routing table. This enabled each node to discover neighbors of a neighbor node or nodes at multi hop distance. To reduce communication cost only updated routing table information was broadcasted. Status of routing tables after routing table broadcast process is given in Figure 7.

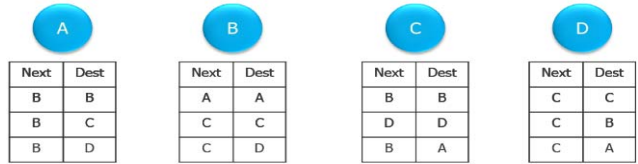

Figure 7. Status of routing tables after routing table broadcast process.

To verify multi hop communication, an application running on node A sent a data transmission request for node D to App Data Manager which in turn communicated with Routing Manger. Routing Manager at node A accessed routing table and selected node B for transmission of data to destination node D. A packet created by Routing Manager was sent to Data Transfer Manager which transmitted it to node B.

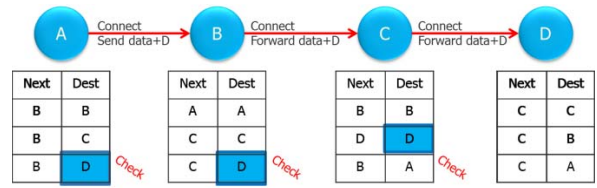

Figure 8. Process of forwarding a data from a source to destination node.

Data Transfer Manager at Node B received the packet and delivered it to a Routing Manager on node B. Routing Manager then selected node C as a next node and performed the same process as on node A. Node C processed the packet and finally delivered it to destination node D using the same process. Packet forwarding process is depicted in Figure 8 whereas packet processing process involving routing layer components is shown in Figure 9.

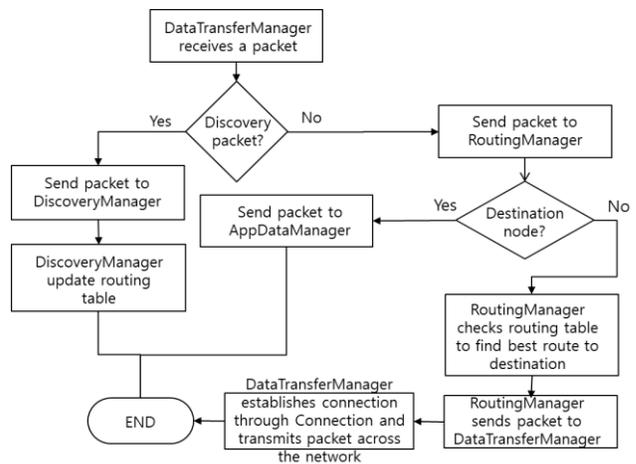

Figure 9. Packet processing diagram.

## VI. Conclusion

Wi-Fi Direct is a new technology that enables mobile devices to directly communicate with each other without a wireless access point. Compare to existing technologies it provides data rates of up to 250 Mbps which is sufficient for several mobile data intensive applications. Wi-Fi Direct technology, however, has two main limitations. It does not support communication between two client devices in a group and it also does not provide support for multi hop communication between nodes across a group which is a key requirement of several IoT and mobile cyber physical system applications. To overcome the limitations, Wi-Fi Direct based multi hop mobile ad hoc network has been developed. The proposed system can be easily extended to support communication between devices using various communication technologies such as ZigBee and Bluetooth. This would enable numerous IoT applications where devices with heterogeneous communication technologies are integrated in to a single network. The proposed system also aims to provide a communication infrastructure for mobile ad hoc cloud systems.

## Acknowledgment

This work was supported by Hankuk University of Foreign Studies Research Fund of 2017 and National Research Foundation Korea.

## References

[1] Shah S. C., Recent Advances in Mobile Grid and Cloud Computing, Intelligent Automation and Soft computing, 2 Feb 2017.

[2] Shah S. C., Energy efficient and robust allocation of interdependent tasks on mobile ad hoc computational grid. Concurrency and Computation: Practice and Experience, 27(5), 2016.

[3] Wi-Fi Peer-to-Peer (P2P) Technical Specification, Wi-Fi Alliance, P2P Task Group, 2014.

[4] Wi-Fi Direct, http://www.wi-fi.org/discover-wi-fi/wi-fi-direct, accessed on 17 Feb 2017.

[5] Marco Conti, Franca Delmastro, Giovanni Minutiello, Experimenting opportunistic networks with Wi-Fi Direct, IFIP Wireless Days, January 2014

[6] Daniel Camps-Mur, Andres Garcia-Saavedra, Pablo Serrano, Device-To-Device Communications with Wi-Fi Direct: Overview and Experimentation, IEEE Wireless Communications, 20(2), June 2013.

[7] Yufeng Duan, Carlo Borgiattino, Claudio Casetti, Carla Fabiana Chiasserini, Paolo Giaccone, Marco Ricca, Fabio Malabocchia, Maura Turolla, Wi-Fi Direct Multi-group Data Dissemination for Public Safety, World Telecommunications Congress, June 2014.

[8] Woo-Sung Jung, Hyochun Ahn, Young-Bae KO, Designing Content-Centric Multi-hop Networking over Wi-Fi Direct on Smartphones, Wireless Communications and Networking Conference, Nov 2014.

[9] Claudio Casetti, Carla Fabiana Chiasserini, Luciano Curto Pelle, Carolina Del Valle, Yufeng Duan, Paolo Giaccone, Content-centric Routing in Wi-Fi Direct Multi-group Networks, IEEE International Symposium on Mobile and Multimedia Networks, July 2015.

[10] Colin Funai, Cristiano Tapparello, Wendi Heinzelman, Supporting Multi-hop Device-to-Device Networks Through WiFi Direct Multi-group Networking, Department of Electrical and Computer Engineering, University of Rochester, December 2015.

[11] Marco Di Felice, Luca Bedogni, Luciano Bononi. The Emergency Direct Mobile App: Safety Message Dissemination over a Multi-Group Network of Smartphones using Wi-Fi Direct, 14th ACM International Symposium on Mobility Management and Wireless Access, 2016